\documentclass[fleqn,twoside]{article}
\usepackage{espcrc2}

\usepackage{graphicx}
\usepackage[figuresright]{rotating}

\newcommand{\AmS}{{\protect\the\textfont2
  A\kern-.1667em\lower.5ex\hbox{M}\kern-.125emS}}

\title{Precise predictions for Higgs production in models with color-octet scalars}

\author{R. Boughezal\address[MCSD]{Institute for Theoretical Physics, University of Zurich\\ Zurich 8057, Switzerland } and
F. Petriello\address[MCSD]{Department of Physics, University of Wisconsin \\ 
        Madison, Wisconsin 53706 USA}%
        \thanks{Work supported by the Swiss National Science Foundation under contract 200020-116756/2 and by the U.S. Department of Energy, Division of High Energy Physics, under contract DE-AC02-06CH11357 and the grant DE-FG02-95ER40896.},
        }
       
\begin{document}

\begin{abstract}
We describe an effective-theory computation of the next-to-next-to-leading order (NNLO) QCD corrections to the gluon-fusion production of a Higgs boson in models with massive color-octet scalars in the ${\bf (8,1)_0}$ representation.  Numerical results are presented for both the Tevatron and the LHC.  The estimated theoretical uncertainty is greatly reduced by the inclusion of the NNLO corrections.  Color-octet scalars can increase the Standard Model rate by more than a factor of two in allowed regions of parameter space.
\vspace{1pc}
\end{abstract}

\maketitle

\section{Introduction}

The hunt for the Higgs boson to uncover its role in electroweak symmetry breaking is now being undertaken at both the Tevatron and the Large Hadron Collider (LHC).  The CDF and D0 collaborations at the Tevatron have recently announced a 95\% exclusion limit on a SM Higgs boson with a mass in the range 
162 GeV $\leq m_h \leq$ 166 GeV~\cite{Aaltonen:2010yv}.  This search requires precise theoretical predictions for the cross section for Higgs production.   The dominant hadronic production mechanism, gluon fusion through a top-quark loop, is known exactly through next-to-leading order in perturbative QCD~\cite{Djouadi:1991tka,Spira:1995rr}.   In the effective theory with $m_t \to \infty$, both the NLO corrections~\cite{Dawson:1990zj} and the NNLO corrections are known~\cite{Harlander:2002wh,Anastasiou:2002yz,Ravindran:2003um}.  When normalized to the full $m_t$-dependent leading-order result, this effective theory reproduces the exact NLO result to better than 1\% for $m_h < 2m_t$ and to 10\% or better for Higgs boson masses up to 1 TeV~\cite{ztalks}.  The QCD radiative corrections drastically alter the Higgs production cross section prediction; for example, the gluon fusion cross section is increased by roughly a factor of three above the LO prediction at the Tevatron after the NNLO corrections are included.  Only at NNLO is an accurate prediction free from debilitating uncertainties obtained.  After inclusion of the NNLO corrections, the remaining uncertainty form uncalculated higher-order terms is estimated to be roughly $\pm 10\%$.  Updated cross sections for Higgs production in gluon fusion for use at the Tevatron and LHC are available in Refs.~\cite{Anastasiou:2008tj,deFlorian:2009hc}.  For a recent review of the status of theoretical predictions for Higgs boson production in the SM, see Ref.~\cite{Boughezal:2009fw}.

The properties of the Higgs boson can be modified  in theories with additional particles, and measurement of these properties consequently serves as a window into physics beyond the SM.  As an example of such a Standard Model extension, we consider new particles transforming in the adjoint representation under the color gauge group.  Several interesting extensions of the SM introduce such states.   Scalars in the $(\bf{8},\bf{1})_{0}$ representation can arise in theories with universal extra dimensions~\cite{Dobrescu:2007xf,Dobrescu:2007yp}.   The primary decay for such states is expected to be into either $b\bar{b}$ or $t\bar{t}$, depending on the scalar mass and other model parameters.  The Tevatron experiments can search for these states via pair production of $(\bf{8},\bf{1})_{0}$ scalars, leading to a four $b$-jet final state.  The search reach in the scalar mass was recently estimated to be 280 GeV~\cite{Dobrescu:2007yp}.  The direct search for these scalars is rendered difficult by the large QCD background.  Reduction of the background requires significant cuts that reduce the signal and therefore the search reach.  It is possible that indirect searches for these scalars, such as via their influence on the Higgs production cross section, can probe masses competitive with direct searches.

We summarize here our recent calculation of the NNLO corrections to the Higgs boson production cross section in models with a $(\bf{8},\bf{1})_{0}$ scalar~\cite{Boughezal:2010ry}.  To be as independent as possible from the origin of this scalar, we study a simple model that couples this scalar to both QCD and the Higgs doublet via renormalizable operators.  We utilize the effective-theory approach valid when both the SM top quark and the new scalar are heavier than the Higgs boson. As a byproduct of our computation, we derive all 
renormalization constants needed to study higher-order corrections in models with these scalars.  We use the renormalization-group equations governing the evolution of the scalar-sector couplings to determine the likely ranges of the various parameters which appear.  

We also study the phenomenological implications of the color-octet scalar for the Higgs production cross section at both the Tevatron and the LHC.  Only at NNLO is the scale dependence sufficiently reduced to allow precise predictions for the scalar-induced effects to be obtained.  The deviations from the Standard Model prediction for Higgs production are large at both colliders.   Deviations of a factor of two, much larger than the estimated uncertainties from higher-order corrections and parton distribution function errors, are obtained for scalar masses near the estimated direct search reach of $m_S \approx 300$ GeV.

\section{Details of the Model}
\label{sec:model}

We begin with the following Lagrangian, which describes the Standard Model coupled to a color-octet scalar in the $(\bf{8},\bf{1})_{0}$ representation:
\begin{eqnarray}
\label{eq:Lfull}
{\cal L}^{full} &=& {\cal L}_{SM}
         + {Tr}\left[D_{\mu} S D^{\mu} S\right]  
         - m_S^{'2} \,{Tr} \left[ S^2 \right] 
\nonumber \\
        && - g_s^2 \,G_{4S} \,{Tr} \left[ S^2 \right]^2-  \lambda_1 H^{\dagger}H \,{Tr} \left[ S^2 \right] 
\nonumber \\
        && -\lambda_h \left(H^{\dagger}H - \frac{v^2}{2}\right)^2.
\end{eqnarray}
$S$ denotes the matrix-valued scalar field $S = S^A T^A$, $H$ indicates the Higgs doublet before electroweak symmetry breaking, $v$ is the Higgs vacuum-expectation value, and $D_{\mu}$ is the covariant derivative for adjoint fields.  The Higgs quartic coupling has been explicitly included to define its normalization.  After electroweak symmetry breaking, the Higgs doublet is expanded as $H = \left(0,(v+h)/\sqrt{2} \right)$ in the unitary gauge.  The mass of the color-octet scalar becomes $m_S^2 = m_S^{'2} + \lambda_1 v^2/2$.  The Feynman rules which describe the scalar couplings to the Higgs boson $h$ and to gluons are easily obtained from Eq.~(\ref{eq:Lfull}).  The free parameters which govern the scalar properties are $m_S$, $\lambda_1$, and $G_{4S}$.  We note that higher-order operators which break the $S \to -S$ symmetry present above and which allow the scalar to decay are obtained in explicit models which contain this state~\cite{Dobrescu:2007xf,Dobrescu:2007yp}.  We neglect them here since we anticipate that they have little effect on the $gg \to h$ production cross section.  We comment on the appearance of the ${Tr} \left[ S^2 \right]^2$ term in the scalar potential.  A quartic-scalar coupling is generated by QCD interactions even if it is set to zero at tree-level.  At NNLO the quartic coupling must be included to obtain a renormalizable result.  We include this operator in the tree-level Lagrangian with a coefficient scaled by $g_s^2$, the QCD coupling constant squared, to permit an easier power-counting of loops.

\section{Calculational details and analytic results}
\label{calc}

\begin{figure}[htbp]
\centerline{
   \includegraphics[width=0.45\textwidth,angle=0]{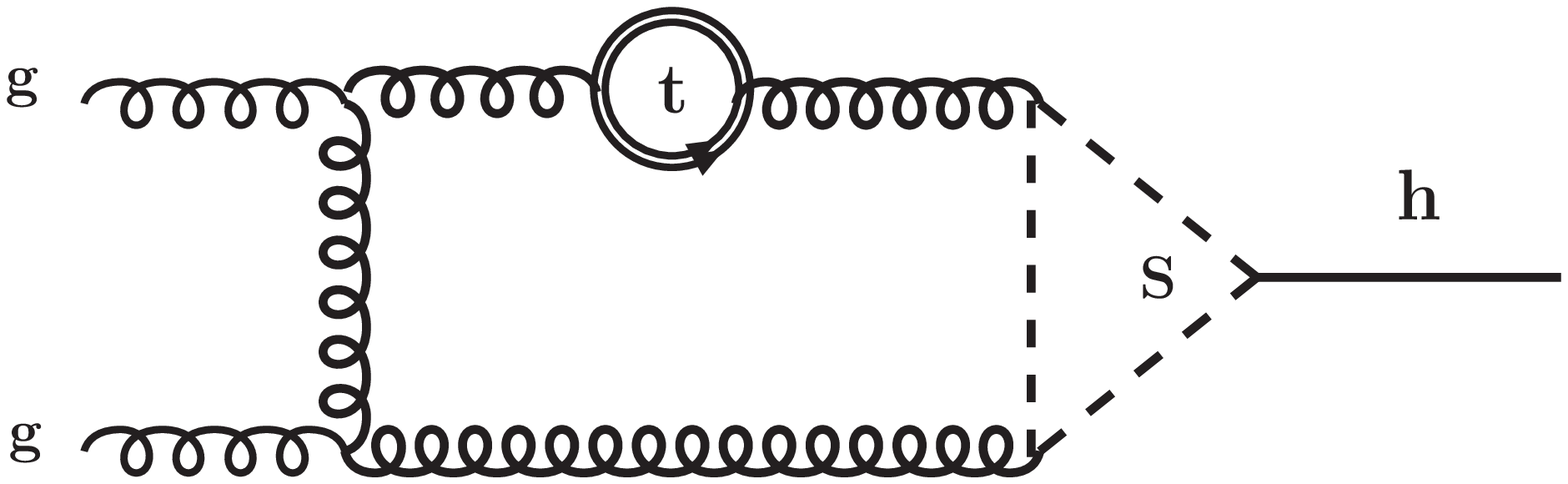}}
\centerline{
    \includegraphics[width=0.45\textwidth,angle=0]{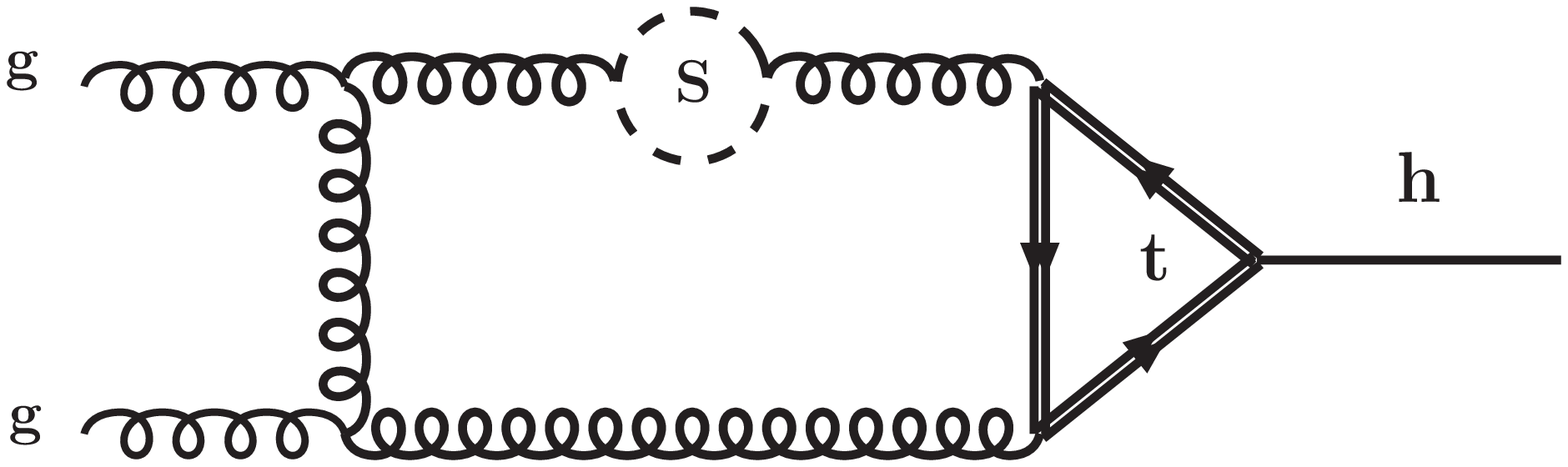}}
   \caption{Example two-scale diagrams contributing to the Wilson coefficient at NNLO.}
   \label{twoscale}
\end{figure}

We sketch here our derivation of the effective Lagrangian describing the interaction of the Higgs boson with gluons through NNLO in the QCD coupling constant.  When both the adjoint scalar and the top quark are heavier than the Higgs boson, they can be integrated out to derive the following effective Lagrangian:
\begin{equation}
\label{eq:Leff}
{\cal L}^{eff} = {\cal L}_{QCD}^{n_l,eff}
         - C_1 \,\frac{H}{v} \,{\cal O}_1,
\end{equation}
where $C_1$ is a Wilson coefficient and the operator ${\cal O}_1$ is
\begin{equation}
{\cal O}_1 = \frac{1}{4} \,{G'}^a_{\mu\nu} {G'}^{a\mu\nu} \, .
\end{equation}
The effective Lagrangian for the gluons and light quarks, ${\cal L}_{QCD}^{n_l,eff}$, has the same form as ${\cal L}_{QCD}^{n_l}$ except that the fields and parameters it contains are rescaled by decoupling constants that account for the effects of the heavy states. In order to distinguish the fields and parameters occurring in the full Lagrangian from those in the effective Lagrangian, we denote the latter ones with a prime.  In this manuscript we integrate out the top quark and the color-octet scalar in a  single step.  Our calculation is therefore a two-scale problem.  This is clearly demonstrated by the example diagrams contributing to the NNLO Wilson coefficient shown in Fig.~\ref{twoscale}.

The Wilson coefficient $C_1$ can be constructed by computing the amplitude for $gg \to h$ in the limit that the initial gluon momenta vanish, as reviewed in Ref.~\cite{Steinhauser:2002rq}.  We start from the relation
\begin{eqnarray}
\label{eq:C01}
\frac{\zeta^0_3 C^0_1}{v}&=&
\frac{\delta^{a_1 a_2}
 \left( g^{\mu_1\mu_2} (p_1\cdot p_2) - p_1^{\mu_2} p_2^{\mu_1} \right)}
     { (N^2-1) (d-2) {(p_1\cdot p_2)^2} } \nonumber \\
   &\times&  {\cal M}^{0,a_1 a_2}_{\mu_1\mu_2}(p_1, p_2)
     |_{p_1 = p_2 = 0}
\end{eqnarray}
between the bare Wilson coefficient $C^0_1$ and the bare amplitude for the process $g g \to H$ in the full theory. Here, $p_1$ and $p_2$ are the momenta of the two gluons, $N$ is the number of colours, and $d=4-2\epsilon$ is the dimension of space-time. The factor $\zeta^0_3$ is the bare decoupling coefficient by which the bare gluon fields in the full and effective theories are related:
\begin{equation}
{G'}^{0,a}_{\mu} = \sqrt{\zeta^0_3} \,{G}^{0,a}_{\mu} \, .
\end{equation}
It can be expressed as
\begin{equation}
\label{eq:zeta03Pi}
\zeta^0_3 = 1 + \Pi^0_G(p=0) \, ,
\end{equation}
where $\Pi^0_G$ is the transverse part of the bare gluon self-energy in the full theory. In order to obtain the Wilson coefficient through NNLO in the QCD coupling constant, ${\cal M}^{0,a_1 a_2}_{\mu_1\mu_2}$ is needed up to three loops, while $\Pi^0_G(p=0)$ is needed through two loops.  We can set all scaleless integrals to zero in dimensional regularization, and as a result, only diagrams containing at least one massive scalar propagator contribute to $\Pi^0_G(p=0)$ and to the right hand side of Eq.~(\ref{eq:C01}).  We perform our calculations in a covariant gauge with gauge parameter $\xi$, leading to the following gluon propagator:
\begin{equation}
\frac{i}{q^2} \left( -g^{\mu\nu} + \xi \frac{q^{\mu}q^{\nu}}{q^2} \right) \, .
\end{equation}
In all diagrams, terms up to first order in $\xi$ are retained.  All $\xi$-dependent terms cancel in $\Pi^0_G(p=0)$ up to two-loop order, and in ${\cal M}^{0,a_1 a_2}_{\mu_1\mu_2}$ up to three-loop order, providing a strong calculational check on the result for $C^0_1$.

We now outline the steps involved in deriving the Wilson coefficient describing the color-octet interactions with the Higgs boson.  Initially, all quantities appearing in Eq.~(\ref{eq:C01}) are expressed in terms of the bare masses $m_S^0$, $m_t^0$, and the bare coupling constants $g_s^0$, $G_{4S}^0$.  We renormalize the scalae and top-quark masses in the   
$\overline{MS}$ scheme according to
\begin{equation}
\label{eq:mSren}
m_S^0 = \sqrt{Z_{m_S}}\, m_S,  \;\;m_t^0 = Z_{m_t}\, m_t.
\end{equation}
The explicit expressions for these renormalization constants, and all others, can be found in Ref.~\cite{Boughezal:2010ry}.  The quartic coupling first appears in the $gg \to h$ amplitude at the two-loop level, and its renormalization is therefore only required to one loop:
\begin{equation}
\label{eq:G4Sren}
G^0_{4S} = Z_{4S}\, G_{4S}\,.
\end{equation}
We convert the bare strong coupling constant of the full theory into the bare coupling of the effective theory using decoupling constants obtained from the ghost self energy and the ghost-gluon vertex up to two-loops, as described in detail in~\cite{Chetyrkin:1997un,Steinhauser:2002rq}.  We then renormalize it in the effective theory using the $\overline{MS}$ scheme:
\begin{equation}
g'^0_s = \mu^{\epsilon}\, Z'_g\, g'_s \, .
\end{equation}
The Wilson coefficient itself requires a renormalization factor~\cite{Spiridonov:1984br,Spiridonov:1988md,Chetyrkin:1996ke}:
\begin{equation}
C_1 = \frac{1}{Z_{11}} \,C^0_1\,.
\end{equation}
Our final result for the renormalized Wilson coefficient is rather lengthy, and can be found in Ref.~\cite{Boughezal:2010ry}. 

\section{Numerical Results}
\label{sec:num}

We now present numerical results for the Higgs boson production cross section in gluon fusion, to study the deviations induced by the color-octet scalar.  We include the effects of the top quark, the scalar, and also the bottom quark on the Higgs cross section.  We comment first on precisely what terms we include in the cross section.  We denote by $\sigma_{T+S}$ the terms obtained by squaring together the top and scalar amplitudes, and keeping both the interference term and the pieces from each separate particle squared.  We let $\sigma_{TB},\sigma_{SB}$ denote the interferences between the bottom-quark amplitude with the top and the scalar pieces, respectively.  $\sigma_{BB}$ indicates the contribution from the bottom-quark amplitude squared.  For the cross section at the $n$-th order in perturbation theory, we 
use the following expression:
\begin{eqnarray}
\sigma^n &=& \sigma_{T+S}^{LO}(m_t,m_S) \, K^n_{EFT}+\sigma_{SB}^{LO}(m_S,m_b)\nonumber \\ &+&\sigma_{BB}^{LO}(m_b) +\sigma_{TB}^{LO}(m_t,m_b).
\end{eqnarray}
$K^n_{EFT}$ denotes the ratio of the $n$-th order cross section over the LO result, with both computed in the effective theory defined in Eq.~(\ref{eq:Leff}).  The cross section multiplying the $K$-factor is the LO cross section with the exact dependence on the scalar and top-quark masses.  The remaining terms account for the scalar-bottom interference, the top-bottom interference, and the bottom-squared contribution at LO with their exact mass dependences.  In the SM, the scaling of the exact LO cross section by the EFT $K$-factor is known to furnish an approximation accurate to the few-percent level or better for Higgs masses below roughly 400 GeV~\cite{ztalks}.  For the scalar, this approximation has been studied against the exact NLO calculation~\cite{Bonciani:2007ex}, and is again accurate to the $1-2$\% percent level for $m_h \leq m_S$, which is the region we focus on here.  If desired, the exact NLO corrections to the top-bottom interference and bottom-squared 
terms~\cite{Spira:1995rr,Anastasiou:2009kn} can be included, as can those to the bottom-scalar interference~\cite{Bonciani:2007ex}.  These affect the cross section at the $1-2\%$ level, and for simplicity are neglected.  Various electroweak corrections known for the SM contribution~\cite{Aglietti:2004nj,Actis:2008ug,Anastasiou:2008tj,Keung:2009bs} are not known for the scalar, and for consistency are neglected.

\begin{figure}[htbp]
   \centering
   \includegraphics[width=0.32\textwidth,angle=90]{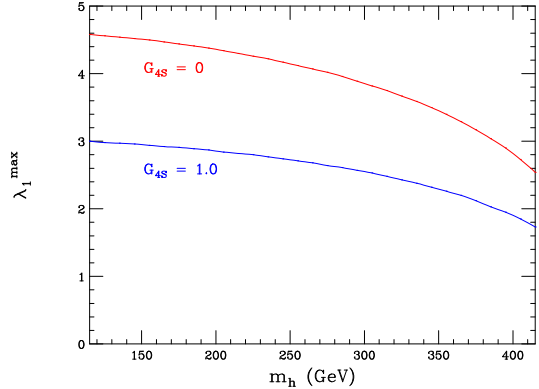}
    \vspace{-0.5cm}
   \caption{Maximum value of $\lambda_1$ allowed by perturbativity for the choices $G_{4S}(v)=1$ and $G_{4S}(v)=0$, as a function of Higgs mass.}
   \label{lam1max}
\end{figure}

We use the pole mass $m_t = 173.1$ GeV for the top quark~\cite{:2009ec}, while for the bottom quark we use the $\overline{MS}$ mass~\cite{Kuhn:2007vp}.  The choice of pole or $\overline{MS}$ mass for the $b$-quark has an insignificant effect on the
fractional deviation between the scalar-induced cross section and the
SM result.  We use the MSTW parton distribution functions~\cite{Martin:2009iq} extracted to the appropriate order in perturbation theory.  For the scalar sector, we must set the parameters $\lambda_1$, $G_{4S}$, and $m_S$.  We perform a renormalization-group analysis to constrain the possible values of the scalar couplings.  We show in Fig.~\ref{lam1max} the maximum allowed value of $\lambda_1(v)$ obtained by demanding $\lambda_1(\mu)$ remain perturbative until $\mu=10$ TeV, as a function of the Higgs mass, for the choices $G_{4S}(v)=1$ and $G_{4S}(v)=0$.  For a complete description of the renormalization-group analysis, we refer the reader to Ref.~\cite{Boughezal:2010ry}.

\section{Results for the Tevatron and the LHC}

\begin{figure}[ht]
   \centering
   \includegraphics[width=0.32\textwidth,angle=90]{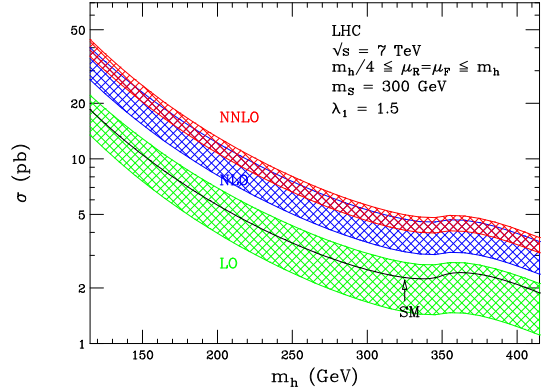}
   \vspace{-0.5cm}
   \caption{Higgs production cross section in gluon-fusion at the LHC for $m_S=300$ GeV as a function of the Higgs boson mass.  The bands indicate the scale variation $m_h/4 \leq \mu \leq m_h$.  From bottom to top, the bands indicate the variations of the LO, NLO, and NNLO cross sections.  All other parameters are as described in the text.  For orientation, the SM result at NNLO for the central value $\mu=m_h /2$ is shown.  We note that the residual scale uncertainty of the SM cross section is approximately $\pm 10\%$.}
   \label{fig:mhplot}
\end{figure}

We now present numerical results for both the Tevatron with $\sqrt{s}=1.96$ TeV and the LHC, for which we assume $\sqrt{s}=7$ TeV.  We begin by showing in Fig.~\ref{fig:mhplot} the LHC cross section for $m_S=300$ GeV at LO, NLO and NNLO in QCD perturbation theory, to see the effect of including higher-order QCD corrections.  The renormalization and factorization scales are equated to $\mu_F = \mu_R = \mu$, and are varied in the range $m_h/4 \leq \mu \leq m_h$, consistent with previous studies of the Higgs production cross section in the SM.  The scale-variation errors are large at both LO and NLO, and the corresponding error bands do not overlap.  Only at NNLO can a reliable prediction for the cross section be made.  It is also clear from this figure that large variations from the SM prediction are possible for scalar masses near the expected Tevatron limit of $m_S \approx 300$ GeV.  The cross section differs from the SM result by more than a factor of two for this parameter value.

To study further the effect of the color-octet scalar on the Higgs cross-section prediction at the Tevatron, we show below in Fig.~\ref{fig:TEVmsplot} the cross sections for the example Higgs mass $m_h =165$ GeV as functions of the scalar mass.  Deviations from the SM are visible over scale errors for scalar masses of up to 1 TeV.   The scalar contributions to the Higgs production cross section are large, and direct searches are hindered by the need to pair produce the scalars and by the large QCD background.  This suggests that the indirect constraints on the scalar parameter space coming from the Tevatron Higgs exclusion limit could be as strong as the direct search reach.

\begin{figure}[ht]
   \centering
   \includegraphics[width=0.32\textwidth,angle=90]{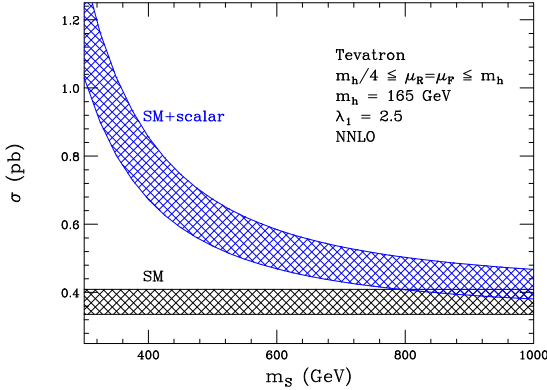}
   \vspace{-0.5cm}
   \caption{Higgs production cross section in gluon-fusion at the Tevatron for $m_h=165$ GeV as a function of the scalar mass.  The bands indicate the scale variation $m_h/4 \leq \mu \leq m_h$.  All other parameters are as described in the text.  Also shown is the SM cross section with its corresponding scale uncertainty.}
   \label{fig:TEVmsplot}
\end{figure}

\section{Conclusions}
\label{sec:conc}

We have discussed a study of the effects of ${\bf (8,1)_0}$ scalars on the Higgs-gluon effective Lagrangian.  We have presented phenomenological predictions for both the Tevatron and LHC.  The scalar-induced effects are large at both colliders, and are visible over both scale and PDF errors for scalar masses of up to 1 TeV.  This suggests that the scalar parameters can be stringently constrained using the exclusion limit on the Higgs boson production cross section obtained by the Tevatron, and that the limits would be competitive with those obtained from direct searches for scalar pair production.  Models that include color-octet scalars are an interesting and phenomenologically-rich example of physics beyond the SM.  The NNLO calculation presented here quantifies precisely the effect of a ${\bf (8,1)_{0}}$ scalar on the Higgs production cross section in gluon fusion, and our phenomenological study shows that the deviations induced by the scalar are large at the Tevatron and the LHC.  The exclusion limit on Higgs production set by the Tevatron collaborations allows for the scalar parameters 
to be stringently constrained.

\end{document}